\documentclass[11pt]{article}
\usepackage{graphicx}% Include figure files
\usepackage{epsfig}

\begin{document}
%\noindent
%\begin{center}
\title{Dynamics of entanglement in a two-mode nonlinear Jaynes-Cummings
 model}
\vspace{1cm}
\author{A. B. M. Ahmed$^{1,a}$ and S. Sivakumar$^{2,b}$\\
$^1$Department of Physics, Indian Institute of Technology Madras \\
  Chennai 600 036, India \\
$^{2}$Materials Physics Division\\
Indira Gandhi Centre for Atomic 
Research  Kalpakkam 603 102, India\\
$^{a}$ abmahmed@physics.iitm.ac.in\\
$^{b}$ siva@igcar.gov.in
}
\maketitle
\vspace{1cm}
\begin{abstract}
         Dynamics of entanglement due to intensity-dependent interaction 
between a two-level atom and a single-mode electromagnetic field in a Kerr 
medium is studied. The form of the interaction is such that the Hamiltonian
 evolution is exactly solvable. The Hamiltonian is shown to be a deformed 
Jaynes-Cummings model admitting a closed, symmetric algebra. Dynamics of 
population inversion and atom-field entanglement are studied taking the 
initial state of the field to be either a coherent state or a squeezed 
vacuum. Analysis is extended to the case of a two-mode cavity field 
interacting with a two-level atom. For the two-mode case, the initial field
 is a pair coherent state or a two-mode squeezed vacuum. Effects due to 
nonlinearity, intensity-dependent interaction and detuning on the dynamics 
are discussed and compared with those of the single-mode case.\\\\
PACS: 42.50.Ct, 42.50.Dv, 03.67.Mn\\
\end{abstract}
\newpage
\noindent
{\large\bf I. Introduction}\\

          Quantum theory allows for correlations that are not possible in any 
classical stochastic process. The source of such correlations is entanglement 
\cite{schrodinger}. Any system with two or more degrees of freedom has the 
possibility of being in an entangled state. Entangled states of atom and 
electromagnetic field in a cavity can be generated in the interaction of a 
two-level system with an external field. A well studied model of atom-field 
interaction is the Jaynes-Cummings (JC) model \cite{jaynes,rempe,shore,chumakov}
, wherein the atom is treated as an electric-dipole interacting with a quantized
 electromagnetic field. The Hamiltonian is 
\begin{equation}
\hat{H}_{JC} = \omega \,\hat{a}^{\dagger} \hat{a} + \frac{1}{2}\nu 
\hat{\sigma}_{z}  + \lambda(\hat{a}^{\dagger}\,\hat{\sigma}_{-} +\hat{a}\,
\hat{\sigma}_{+}),
\end{equation}
where $\lambda$ is the coupling constant, $\omega$ is the field frequency and 
$\nu$ is the atomic transition frequency. The operators $\hat{a}$ and 
$\hat{a}^{\dagger}$ are the annihilation and creation operators of the field; 
$\sigma_{z}$ is the Pauli matrix, $\hat{\sigma}_{+}$ and $\hat{\sigma}_{-}$ are
 the raising and lowering operators of the atomic states. The ground and the 
excited states of the atom are represented by $|g\rangle$ and $|e\rangle$ 
respectively. The action of the relevant operators on these states are 
\begin{eqnarray}
\hat{\sigma}_{z} |e\rangle &=& \vert e\rangle,\,\,\, \hat{\sigma}_{z} 
\vert g\rangle =-\vert g\rangle,\nonumber\\
\hat{\sigma}_{+} |g\rangle &=& |e\rangle, \,\,\, \hat{\sigma}_{+} |e\rangle = 0,
\nonumber\\
\hat{\sigma}_{-} |e\rangle &=& |g\rangle, \,\,\, \hat{\sigma}_{-} |g\rangle = 0.
\nonumber
\end{eqnarray}
 The first two terms in the Hamiltonian $\hat{H}_{JC}$ correspond to the 
energies of the field and the atom respectively. The last term accounts for the
atom-field interaction. This model has been used extensively in studying 
various quantum optical systems. A generalization of this model is useful in 
analysing position-dependent interaction strength \cite{buck, vogel, dematos}.
 To account for such a dependence, the interaction term is modified to 
\begin{equation}
\hat{H}_{int} = \lambda(f(\hat{a}^{\dagger} \hat{a}) \hat{a} \hat{\sigma}_{+} 
+ \hbox{adjoint}) .
\end{equation}
The interaction term in the JC model is obtained when $f(\hat{a}^{\dagger} 
\hat{a})=\hat{I}$, the identity operator. With any other form for 
$f(\hat{a}^{\dagger} \hat{a})$, the model is referred as Nonlinear 
Jaynes-Cummings(NJC) model. The operator-valued function $f$ carries 
information about the position-dependent interaction strength. If the field has
 nonlinear dependence on the field amplitude, as in the case of Kerr 
interaction \cite{gora92, joshipuri, werner}, the Hamiltonian is generalized
 further to 
\begin{equation}
\hat{H} = \omega\,\hat{a}^{\dagger} \hat{a} + \frac{1}{2}\nu \hat{\sigma}_{z} 
 + \chi  \hat{a}^{\dagger\,2} \hat{a}^{2} + \lambda(f(\hat{a}^{\dagger} \hat{a})
 \hat{a} \hat{\sigma}_{+} + \hbox{adjoint}).
\end{equation}
This Hamiltonian, with $f(\hat{a}^{\dagger} \hat{a}) = \hat{I} $, models the 
interaction between a field in a Kerr medium and a two-level atom. The 
quadratic term $\hat{a}^{\dagger \, 2} \hat{a}^{2}$ accounts for the Kerr 
effect. We consider a special form of $f(\hat{a}^{\dagger} \hat{a})$, so chosen
 that the model Hamiltonian in Eq. (3) is exactly solvable \cite{sivajopa, 
sivaijtp}. Choosing $f(\hat{a}^{\dagger} \hat{a}) = \sqrt{1+k \hat{a}^{\dagger}
 \hat{a}}$, where $0\leq k \leq 1 $, the Hamiltonian $\hat{H}$ is written as 
\begin{equation}
\hat{H}_{S} = \omega\,\hat{K}^{\dagger} \hat{K} + \frac{1}{2}\nu \,
\hat{\sigma}_{z}  + \lambda(\hat{K}^{\dagger} \hat{\sigma}_{-} + \hbox{adjoint})
,
\end{equation}
where $\chi=k\omega$. The suffix ``S" indicates that the Hamiltonian describes
 the interaction of a single-mode of a cavity field with a two-level atom. The 
operators $\hat{K} = \sqrt{1+k \hat{a}^{\dagger} \hat{a}}\, \hat{a}$ and 
$\hat{K}^{\dagger}=\hat{a}^{\dagger} \,\,\sqrt{1+k \hat{a}^{\dagger} \hat{a}}$ 
are deformed annihilation and creation operators respectively. In this work, 
the parameter $k$ is taken to be nonnegative. Nevertheless, when $k=-1$, the 
operators $\hat{K}$ and $\hat{K}^{\dagger}$ are the well-known 
Holstein-Primakoff realization of spin operators. It is interesting to note 
that when $k=0$, the Hamiltonian is the JC Hamiltonian. Nonzero values of $k$ 
amount to including Kerr effect and intensity-dependent interactions. In 
essence, nonlinearity and intensity-dependent interactions are included 
\textit{via} deformation. The operators $\hat{K}$, $\hat{K}^{\dagger}$ and 
$\hat{K}_{0} = k\hat{a}^{\dagger} \hat{a}+\frac{1}{2} $ satisfy the commutation
 relations,
\begin{equation}
[\hat{K},\hat{K}^{\dagger}] = 2\hat{K}_{0} , \,\,\, [\hat{K}_{0},
\hat{K}^{\dagger}] = k\hat{K}^{\dagger}, \,\,\, [\hat{K}_{0},\hat{K}] = 
-k\hat{K}, 
\end{equation} 
with $[\,\, , \,\,]$ denoting the commutator of the enclosed operators. Hence, 
the symmetric set $ \{\hat{K},\hat{K}^{\dagger} , \hat{K}_{0}  \}$ forms a 
closed algebra under commutation. Further, when $k=0$ the algebra is the 
Heisenberg-Weyl algebra generated by $\{\hat{a}, \hat{a}^{\dagger}, 
\hat{I} \}$. Another interesting limit corresponds to $k=1$. In this case, 
the algebra defined in Eq. (5) is SU(1,1) algebra. Thus, two important algebras,
 namely, the Heisenberg-Weyl and the SU(1,1), are the special cases of the 
generalized algebra defined in Eq. (5). Mathematically, choosing nonzero $k$ 
values amounts to deforming the Heisenberg-Weyl algebra. Thus, the Hamiltonian
 $\hat{H}_{S}$ can be viewed as a deformed JC Hamiltonian. Though there is a 
formal resemblance between $\hat{H}_{JC}$ and $\hat{H}_{S}$, the deformation 
($k \ne 0$) allows to include nonlinearity and intensity-dependent coupling. 
Under suitable limits, the Hamiltonian $\hat{H}_{S}$ approximates many 
well-studied Hamiltonians. For instance, when $k=0$, $\hat{H}_{S}$ becomes 
the usual JC model; when the mean photon number of the field satisfies 
$k\langle \hat{a}^{\dagger}\hat{a}\rangle >> 1$, then 
$\sqrt{1+k\hat{a}^{\dagger}\hat{a}} \sim \sqrt{k\hat{a}^{\dagger}\hat{a}}$ and
 $\hat{H}_{S}$ approximates the Buck-Sukumar Hamiltonian \cite{buck}.\\

    In the single-mode field-atom bipartite system, the field is a continuous 
variable (infinite-dimensional Hilbert space) system while the atom is 
described in a two-dimensional Hilbert space. Recently, much work has been 
done on systems with two continuous variable subsystems; the entangled coherent
 states \cite{wangsanders,joengralph} being the most widely studied. Such 
entangled continuous systems have desirable features which can be exploited in
 teleportation, quantum communication, etc \cite{braunstein, wangsanders,
tombesi,sanders,vanenk,vanenkhirota,joengralph,gerry,chai}. Two-mode 
electromagnetic fields can be entangled by allowing them to interact with a 
two-level atom in a cavity. A basic scheme to perform this was explored in 
\cite{larson} by employing a two-mode JC model. Various versions  of the 
two-mode JC model \cite{pramana} have been used in important contexts like the
 phase damping \cite{obeda08optcomm}, entanglement distribution 
\cite{wangepjd}, entangling field modes with a moving atom \cite{liao}, entropy
 squeezing \cite{abdelatyJoOpticsB} etc. In this work, a natural extension of 
$\hat{H}_{S}$ is studied. One possibility is to consider   
\begin{equation}
\hat{H}_{T} = \omega_{1}\hat{K}_{1}^{\dagger}\hat{K}_{1} 
+ \omega_{2}\hat{K}_{2}^{\dagger}\hat{K}_{2} +\frac{1}{2}\hat{\sigma}_{z} 
+\lambda(\hat{K}_{1}^{\dagger}\hat{K}_{2}^{\dagger}\hat{\sigma}_{-} 
+ \hat{K}_{1}\hat{K}_{2}\hat{\sigma}_{+} ),
\end{equation}
where $\hat{K}_{1} = \sqrt{1+k_{1} \hat{a}^{\dagger} \hat{a}}\, \hat{a}$ and 
$\hat{K}_{2} = \sqrt{1+k_{2} \hat{b}^{\dagger} \hat{b}}\, \hat{b}$. The 
creation operators of the two modes are $\hat{a}$ and $\hat{b}$ respectively;
 the corresponding annihilation operators are $\hat{a}^\dagger$ and 
$\hat{a}^\dagger$. The suffix ``T" indicates that the composite system 
involves a two-mode field. The operators $\hat{K}_{1}$ and $\hat{K}_{2}$ 
correspond to the two field modes of the cavity. The deformation parameters 
$k_{1}$ and $k_{2}$ satisfy $0\leq k_{1},k_{2} \leq 1 $. The operators in 
each of the sets $ \{\hat{K}_{1},\hat{K}_{1}^{\dagger} , \hat{K}_{10}= 
k_{1}\hat{a}^{\dagger} \hat{a}+\frac{1}{2} \}$ and $ \{\hat{K}_{2},
\hat{K}_{2}^{\dagger} , \hat{K}_{20}= k_{2}\hat{b}^{\dagger} \hat{b}
+\frac{1}{2} \}$, satisfy  the closed algebra; defined in Eq. (5). The 
dynamics dictated by the two-mode Hamiltonian $\hat{H}_{T}$ is exactly 
solvable. This Hamiltonian, like its one-mode counterpart $\hat{H}_{S}$, 
approximates, many of the known two-mode JC models. For instance, if $k_{1}
=k_{2}=0$ then $\hat{H}_{T}$ corresponds to the Hamiltonian studied in the 
context of phase damping \cite{obeda08optcomm}. When the mean photon number is 
large enough so that $1+k\langle \hat{a}^{\dagger} \hat{a}\rangle \approx  
k\langle \hat{a}^{\dagger} \hat{a} \rangle$ and $1+k\langle \hat{b}^{\dagger} 
\hat{b}\rangle \approx  k\langle \hat{b}^{\dagger} \hat{b} \rangle$, then the
 interaction term is approximated by $\sqrt{\hat{b}^{\dagger}\hat{b}}\,\, 
\sqrt{\hat{a}^{\dagger}\hat{a}}\,\,\hat{a}\hat{b}\hat{\sigma}_{+} 
+ \hbox{adjoint}$ and the resulting Hamiltonian is the two-mode Buck-Sukumar
 Hamiltonian. \\    

    In the present work, dynamics dictated by the Hamiltonian $\hat{H}_{S}$ 
and $\hat{H}_{T}$ are studied; in particular, the dynamics of population 
inversion and of entanglement are analyzed. This paper is organized in the 
following way. Results on the single-mode system are presented first and those
 on the two-mode system are presented subsequently. In section II, dynamics of
 entanglement in the single mode case is discussed. The effects of nonlinearity
 and intensity-dependent coupling on the dynamics are studied. In section III,
 analyses on two-mode case are presented. Dynamics of various tangles in the 
system are studied, and wherever possible, the results are compared with those
 of the single mode case. \\\\
\noindent
{\large\bf II. Single-mode NJC model}\\

       In the single-mode NJC model, there are two degrees of freedom in the 
system, corresponding to the field and the two-level atom respectively.  The
 Hamiltonian of the system is taken to be $\hat{H}_{S}$. The initial state of 
the atom-field system is taken to be a product state. Entanglement between the
 atom and the field is generated by the interaction. The time evolution of 
the atom-field state is obtained by solving
\begin{equation}
i\frac{\partial }{\partial t}|\psi(t)\rangle = \hat{H}_{S} |\psi(t)\rangle,
\end{equation}           
where $ |\psi(t)\rangle $ represents the state of the atom-field system at 
time `t'. The explicit form of $ |\psi(t)\rangle $ is $\sum_{n=0}^{\infty}\,
[\,C_{e,n} (t)|e,n\rangle + C_{g,n} (t)|g,n\rangle ]$; the state $|e,n\rangle$
 [respectively, $|g,n\rangle $ ] represents the field in the number state 
$|n\rangle$ and the atom in the state $|e\rangle$[respectively, $|g\rangle$]. 
The coefficients $C_{e,n} (t)$ and $C_{g,n}(t) $ are \cite{sivaijtp}
\begin{equation}
%\begin{split}
e^{-i\Delta_{n} t/2} C_{e,n} (t) = [\cos(\frac{\Omega_{n}t}{2}) 
- \frac{i\Delta_{n}}{\Omega_{n}}\sin(\frac{\Omega_{n}t}{2})]C_{e,n}(0)
       -\frac{2i \lambda \eta_{n}}{\Omega_{n}}\,\, 
\sin(\frac{\Omega_{n}t}{2})\,\, C_{g,n+1}(0),
%\end{split}
\end{equation}
\begin{equation}
%\begin{split}
e^{i\Delta_{n}t/2} C_{g,n+1} (t) = [\cos(\frac{\Omega_{n}t}{2}) 
+ \frac{i\Delta_{n}}{\Omega_{n}}\sin(\frac{\Omega_{n}t}{2})]C_{g,n+1}(0)
       -\frac{2i \lambda \eta_{n}}{\Omega_{n}}\,\, 
\sin(\frac{\Omega_{n}t}{2})\,\, C_{e,n}(0),
%\end{split}
\end{equation}
where $\Delta=\nu-\omega$, $\Delta_{n} = \Delta - 2k\omega n$, 
$\Omega_{n} = \sqrt{\Delta_{n}^2 + 4 {\lambda}^{2}\eta_{n}^{2}}$ and 
$\eta_{n} = \sqrt{(1+n)(1+kn)}$. These analytical results have been used to
 arrive at the results presented subsequently.\\
\noindent

      The initial state of the field is taken to be either a coherent state,
\begin{equation}
|\alpha \rangle =  e^{-\frac{|\alpha|^2}{2}}\,\,\sum_{n=0}^{\infty}
\frac{\alpha^{n}}{\sqrt{n!}} |n\rangle,   \,\,\,\,\,\,\,\,\,\,\,\,\,\, 
\alpha \in \it{C},
\end{equation}
or a squeezed vacuum 
\begin{equation}
|z \rangle =  \frac{1}{\sqrt{\cosh r}}\,\sum_{l=0}^{\infty}(-1)^{l}
\frac{\sqrt{(2l)!}}{2^{l}\,l!}(e^{i\theta} \, \tanh r)^{l} |2l\rangle,
 \,\,\,\,\,\, z=re^{i\theta}, \,\, r = |z|. 
\end{equation}
The states considered in this work correspond to $\alpha=\sqrt{30}$ and 
$z=2.402$.  The  values are chosen so that the mean number of photons in the 
CS and that in the SV are equal to 30.  
The initial state of the system is $(\sum_{n} f_{n} |e,n\rangle)$, where the 
coefficients $f_{n}$ of the field state are chosen to be those of 
$|\alpha \rangle $ or $|z \rangle $. Thus, the initial conditions, 
$C_{e,n}(0)$ and $C_{g,n}(0)$, are 
\begin{equation}
C_{e,n}(0) = f_{n};\,\,\,\,  C_{g,n}(0) = 0.
\end{equation}
These are the initial conditions when the atom evolves from its excited state.
 The density operator for the system is the projection operator 
$|\psi(t)\rangle\langle\psi(t)|$, expressed in the $\{|e,n\rangle , |g,n\rangle 
\}$ basis as
\begin{equation}
\hat{\rho} = |\psi(t)\rangle\langle\psi(t)| =\sum_{a,b=e,g}\,\,
\sum_{n,m=0}^{\infty} C_{a,n}(t)C^{\ast}_{b,m}(t)|a,n\rangle \langle b,m|.
\end{equation} \\
\noindent

\begin{figure}
\centering
\includegraphics[height=15cm,width=15cm]{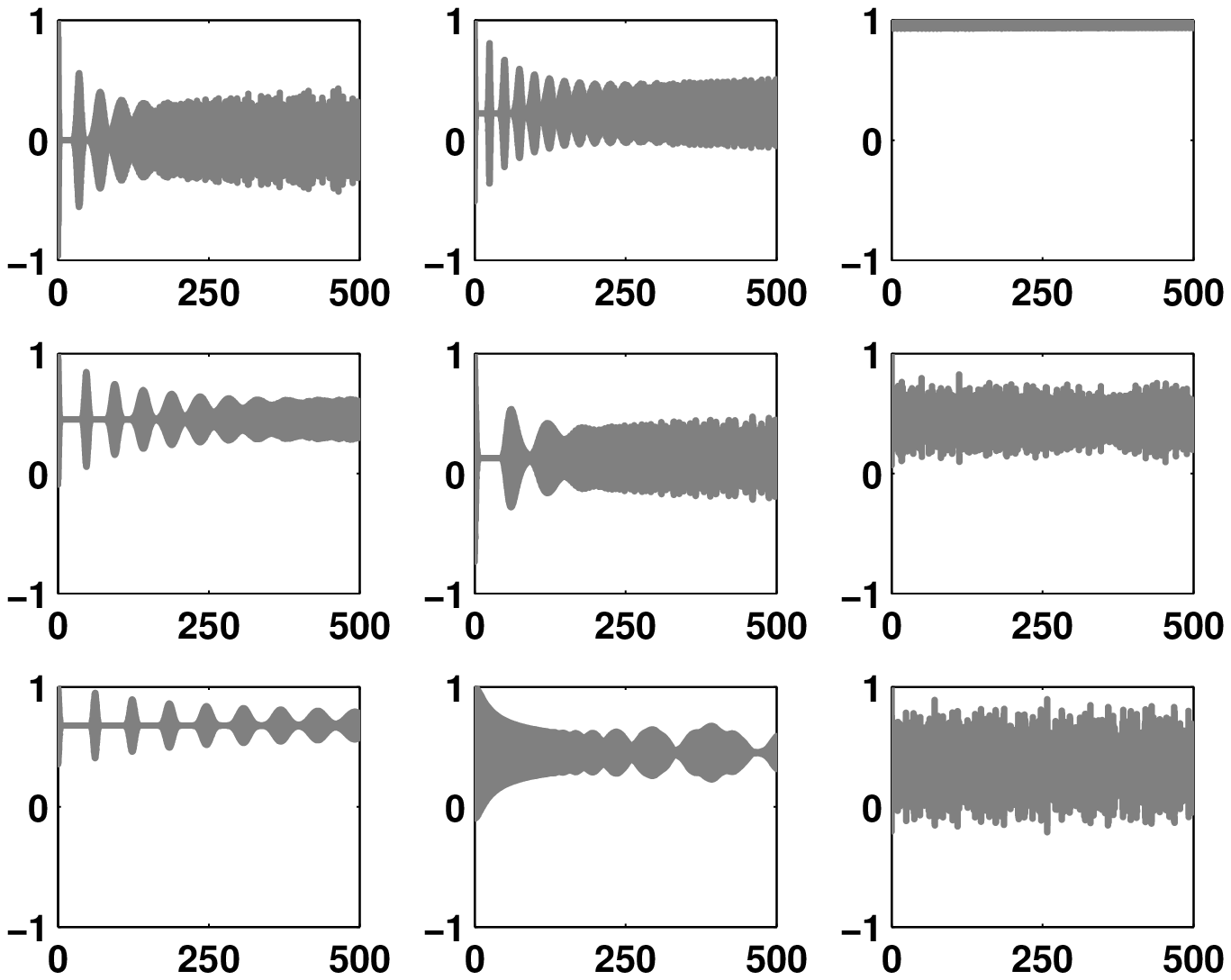}
\caption {{\small Variation of population inversion versus scaled time 
($\lambda t$) when the initial state is $|\alpha \rangle|e\rangle$; the field
 state in the coherent state $|\alpha \rangle$ ($\alpha=\sqrt{30}$) and the 
atom in the excited state $|e\rangle$. The coupling constant is $\lambda = 
0.001$.  Successive columns correspond to $k$ equal to 
$0,\,10^{-4},\,10^{-3}$. In the first and second columns, top most plot 
corresponds to $\Delta=0$ and successive plots correspond to $\Delta = 
0.01(<\Delta_{c}),\,\Delta= 0.016061 (=\Delta_{c})$. In the third column, 
top most plot correspond to $\Delta=0$ and successive plots correspond to 
$\Delta = 0.05(<\Delta_{c}),\,\Delta= 0.061061 (=\Delta_{c})$.}}
\label{fig:pics}
\end{figure}

          Population inversion $W_{S}(t)$ is defined as the difference in the 
probabilities of the atom to be in the states $|e\rangle$ and  $|g\rangle$. 
Hence,
\begin{equation}
W_{S}(t) = Tr[\hat{\rho} \hat{\sigma}_{z}] = \sum_{n=0}^{\infty} 
[|C_{e,n}(t)|^{2}- |C_{g,n}(t)|^{2}  ],
\end{equation}
where Tr stands for tracing over both the atomic and field states. Evolution 
of $W_{S}(t)$ is shown in Fig. \ref{fig:pics}, when the field is initially the
 coherent state $\vert\alpha \rangle$. The figures stacked in the first column
 correspond to $k=0$; the second and third columns correspond to $k=10^{-4}$ 
and  $k=10^{-3}$ respectively.  Different rows of figures give information on
 the effect of detuning. The figures in the first row correspond to 
$\Delta=0$. A special feature of the Hamiltonian $\hat{H}_{S}$ is the 
existence of a minimum for the Rabi frequency $\Omega_{n}$, for some 
$n=\bar{n}$. The numerical value of  $\bar{n}$ is related to the other 
parameters through \cite{sivaijtp}
\begin{equation}
\bar{n} = \frac{k \Delta - {\lambda}^{2}{\omega}^{2}(1+k)}
{2k\omega(k+{\lambda}^{2})}.
\end{equation}     
If the mean value of $\langle \hat{a}^{\dagger}\hat{a}\rangle$ is chosen to 
be $\bar{n}$, the corresponding $\Delta$ satisfying the Eq. (15) is the 
critical detuning $\Delta_{c}$. In the resonant JC model ($k=0,\Delta=0$), 
the perfect collapses of the population inversion are Gaussian modulated 
cosines \cite{eberly,fillipowicz}.  The dynamics of $W_{S}(t)$ is known to 
have distinct features, such as `super structures' in the 
``collapse - revival'' pattern, when $\Delta \approx \Delta_{c}$, 
{\it i.e.}, the envelope of collapse or revival is not a Gaussian modulated 
cosine \cite{gora92}. This is shown in the figure located at the third row of 
the second column in Fig. \ref{fig:pics}.  It is to be noted that a minimum in
 the Rabi frequency is possible only when $k \ne 0$.\\
\noindent

           The first column of figures in Fig. \ref{fig:pics} shows the 
effect of higher detuning on the dynamics of $W_{S}(t)$. Larger detuning 
inhibits the interaction between the atom and the field. This is expected 
based on the Fermi's Golden rule which states that interaction probability 
becomes smaller as the field frequency deviates from the resonant condition.
 Hence, with large detuning the atom evolves such that it has substantial
 overlap with its initial state. The last figure corresponds to  $\Delta = 
0.016$ and the $W_{S}(t)$ profile indicates that the atom practically remains
 in the excited state.\\
\noindent

     Dynamics dictated by $\hat{H}_{S}$ differs from that of $\hat{H}_{JC}$ 
due to the Kerr term and the intensity-dependent interaction in $\hat{H}_{S}$.
 The strength of the nonlinearity($\propto k \hat{a}^{\dagger\,2}\hat{a}^{2}$)
 and that of the coupling($\propto \sqrt{1+k\hat{a}^{\dagger}\hat{a}}$) 
increase   with $k$. It is found that if $k=0.01$, even under resonant 
condition, the state of the atom has substantial overlap with the excited 
state $|e\rangle$. This implies that the presence of cavity field has not 
resulted in any significant change of atomic state. As noted earlier, the 
atom-field interaction becomes insignificant if the interaction is 
non-resonant due to large detuning in the usual JC model, which is the 
undeformed $\hat{H}_{S}$. Thus, the effect of large ``$k$'' mimics the 
effect of large detuning. The origin of this ``inhibition of emission" from 
the excited state is mostly due to the the nonlinear term 
$k  \hat{a}^{\dagger \,2}\hat{a}^{2}$. In the case of coherent states, 
$k  \hat{a}^{\dagger\,2}\hat{a}^{2}$  term contributes 
$|\alpha|^{4}+|\alpha|^{2}$ to the total energy while the contribution from 
the interaction term $\sqrt{1+k\hat{a}^{\dagger}\hat{a}}\hat{a}$ is 
$\sim \sqrt{1+k|\alpha|^{2}}\,\alpha$. For small value of mean photon 
number($\langle \hat{a}^{\dagger}\hat{a}\rangle \sim 1$), the contributions 
of the nonlinear term and the interaction term to the energy are nearly equal.
 However, for large amplitude coherent state, the nonlinear term dominates 
over the interaction. To bring out these features the variation of 
$W_{S}(t)$ is plotted in Fig. \ref{fig:picsspecial} for $|\alpha|^{2}=1$ and 
$|\alpha|^{2}=30$. The interaction is assumed to be resonant. From the figures
 it is seen that as $k$ increases, the quantity $W_{S}(t)$ stays close to 
unity if $|\alpha|^{2}=30$ and oscillates if $|\alpha|^{2}=1$. In the large
 amplitude case, the population inversion remains close to unity while it 
swings between 1 and 0.2 when $|\alpha|^{2} \sim 1$.  The former corresponds
 to dominant nonlinearity while the later corresponds to nearly equal 
magnitudes of the nonlinearity and the interaction.  Thus, the 
``inhibition of emission", despite the presence of a resonant field, is due
 to the nonlinear term. \\
\noindent

     Even at smaller values of $k$ the effect of nonlinearity can be dominant.
 The last column of Fig. \ref{fig:pics} corresponds to $k=10^{-3}$, an order 
of magnitude smaller than $10^{-2}$. The initially excited atom is left 
almost undisturbed in the resonant case as evident from the nearly 
time-independent $W_{S}(t)$. As the detuning increases, the atom begins to
 interact with the field, and, consequently, the oscillations in $W_{S}(t)$
 are larger. This is inferred by comparing the figures in the first row 
(zero detuning) of Fig. \ref{fig:pics} with those in the successive rows 
(higher detunings). These counter-intuitive features, that is, 
noninteracting, resonant field and interacting, non-resonant field, are due
 to the presence of nonlinearity.\\   

\noindent

\begin{figure}
\centering
\includegraphics[height=15cm,width=15cm]{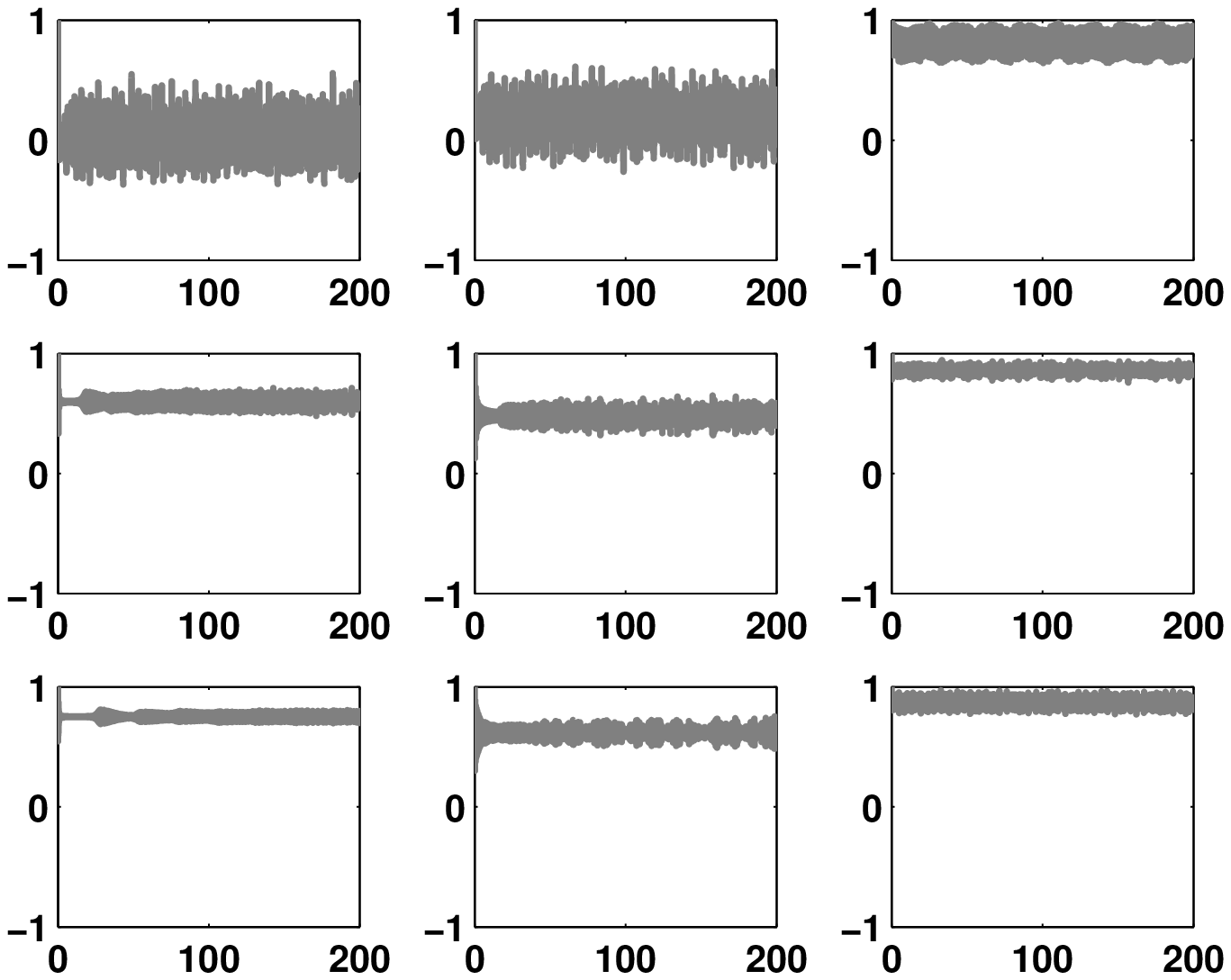}
\caption {{\small Variation of population inversion with scaled time 
($\lambda t$). The initial state of the field is taken to be squeezed vacuum 
state $|z \rangle$ ($z = 2.402$) and that of the atom is the excited state 
$|e\rangle$. Successive columns correspond to $k$ equal to 
$0,\,10^{-4},\,10^{-3}$. In the first and second columns, top most plot 
corresponds to $\Delta=0$ and successive plots  correspond to $\Delta = 
0.01(<\Delta_{c}),\,\Delta= 0.016061 (=\Delta_{c})$. In the third column, top
 most plot corresponds to $\Delta=0$ andn successive plots  correspond to 
$\Delta = 0.05(<\Delta_{c}),\,\Delta= 0.061061 (=\Delta_{c})$. Other 
parameters are the same as in Fig. 1.}}
\label{fig:pisv}
\end{figure}

\begin{figure}
\centering
\includegraphics[height=5cm,width=12cm]{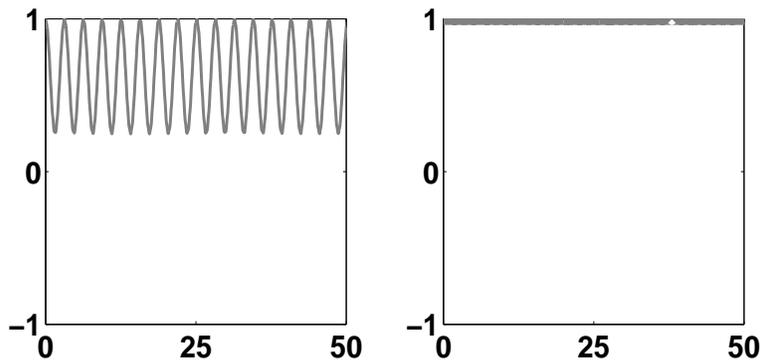}
\caption{ {\small  Temporal profile of population inversion with scaled time 
($\lambda t$). The initial state of the field is coherent state 
$|\alpha\rangle$ and mean number of photons in the mode is 
$|\alpha|^{2}= 1$(left) and $|\alpha|^{2}= 30$(right). The initial state of 
the atom is taken to be excited state $|e\rangle$. The interaction is assumed
 to be resonant ($\Delta=0$). The coupling constant is $\lambda = 0.001$ and 
$k=10^{-2}$.}}
\label{fig:picsspecial}
\end{figure}

            In Fig. \ref{fig:pisv}, the dynamics of $W_{S}(t)$ is shown when
 the field is initially a squeezed vacuum. Some of the features present in 
the interaction of the atom with coherent state are absent. Even though, the
 mean number of photons in the squeezed vacuum is chosen to satisfy Eq. (15), 
there are no ``super" structures in the population inversion. However, as in
 the coherent state case, the field-atom interaction is strengthened at 
nonzero detuning when the value of the deforming parameter $k$ is nonzero but
 small ($~10^{-3}$). The mean number of photons in the CS and the SV are the
 same.  Hence, the differences in the dynamics are due to the different photon
 statistics of the two states.\\

\noindent

\begin{figure}
\centering
\includegraphics[height=15cm,width=15cm]{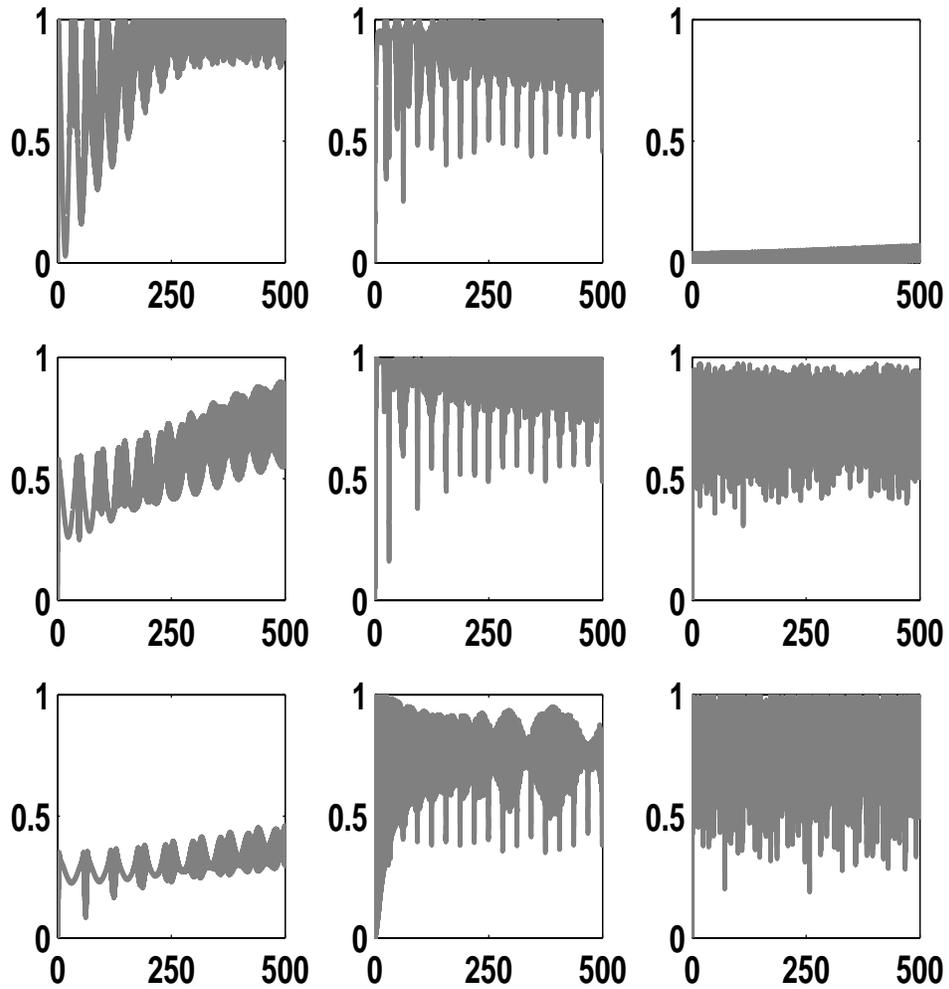}
\caption {{\small Entanglement versus scaled time ($\lambda t$). The initial 
state of the field is the coherent state $|\alpha\rangle$ ($\alpha=\sqrt{30}$) 
and that of the atom is the excited state $|e\rangle$. Successive columns 
correspond to $k$ equal to $0,\,10^{-4},\,10^{-3}$. In the first and second 
columns, top most plot corresponds to $\Delta=0$ and successive plots 
correspond to $\Delta = 0.01(<\Delta_{c}),\,\Delta= 0.016061 (=\Delta_{c})$. 
In the third column, top most plot corresponds to $\Delta=0$ and successive 
plots correspond to $\Delta = 0.05(<\Delta_{c}),\,\Delta= 0.061061 
(=\Delta_{c})$. Other parameters are the same as in Fig. 1.}}
\label{fig:entanglecs}
\end{figure}

\begin{figure}
\centering
\includegraphics[height=15cm,width=15cm]{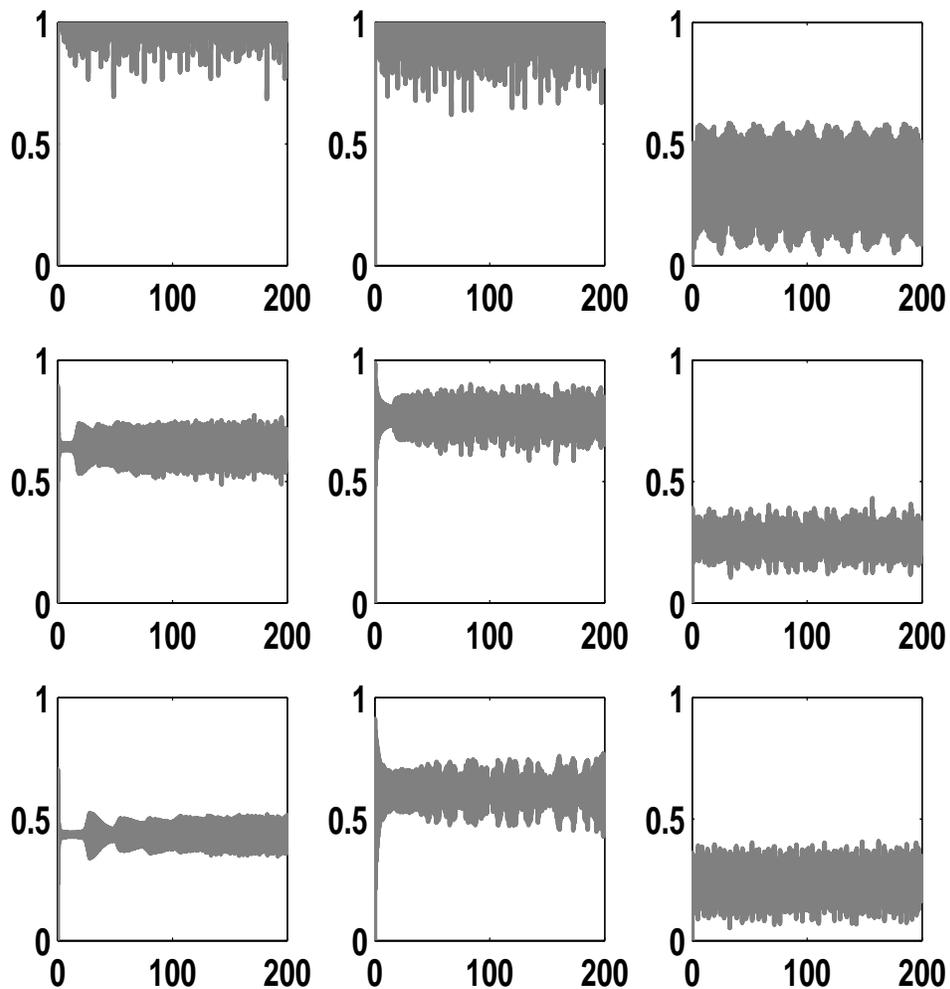}
\caption {{\small Entanglement versus scaled time ($\lambda t$). The initial 
state of the field is the squeezed vacuum state $z\rangle$ ($z=2.402$) and 
that of the atom is the excited state $|e\rangle$. Successive columns 
correspond to $k$ equal to $0,\,10^{-4},\,10^{-3}$. In the first and second 
columns, top  ost plot corresponds to $\delta=0$ and successive plots 
correspond to $\Delta = 0.01(<\Delta_{c}),\,\Delta= 0.016061 (=\Delta_{c})$. 
In the third column, top most plot corresponds to $\Delta=0$ and successive 
plots correspond to $\Delta = 0.05(<\Delta_{c}),\,\Delta= 0.061061 
(=\Delta_{c})$. Other parameters are the same as in Fig. 1.}}
\label{fig:entanglesv}
\end{figure}

         Interaction between the atom and the field entangles them. The 
initial state $\sum f_{n} |e,n\rangle $ is a product state and hence the 
entanglement is zero. To study the effectiveness of the interaction in 
entangling the atom and field system, linear entropy is used as the measure of
 entanglement. If $\hat{\rho}$ is the density operator of the composite 
system, its linear entropy ($L$)  is defined as 
\begin{equation}
L = 2(1 -Tr[\hat{\rho}^{2}_{f}] ) = 2(1 - Tr[\hat{\rho}^{2}_{a}]).
\end{equation}
Here $\hat{\rho}_{f}$ [respectively, $\hat{\rho}_{a} $] is the reduced density
 operator of the field [respectively, atom] obtained by partially tracing over
 the atomic [respectively, field] states. For instance, the reduced density 
operator $\hat{\rho}_{a}$ for the atom is 
\begin{equation}
\hat{\rho}_{a} =\left(
\begin{array}{cc}
\sum_{n}|C_{e,n}(t)|^{2}   & \sum_{n}C^{\ast}_{e,n}(t)C_{g,n}(t) \\
\sum_{n}C_{e,n}(t)C_{g,n}^{\ast}(t) & \sum_{n}|C_{g,n}(t)|^{2}
\end{array}
\right).
\end{equation} 
Even though other measures such as von Neumann entropy exist, in the case of 
pure states all entanglement measures are equivalent. In this work, linear 
entropy is used as the entanglement measure as it is easier to compute.  The 
two quantities, namely, $W_{S}$ and $L$, are related by
\begin{equation}
L = 1 - W_{S}^{2} - 4|\sum C_{e,n}^{\ast}C_{g,n}|^{2}.
\end{equation}
The quantity $|\sum C_{e,n}^{\ast}C_{g,n}|$  is the atomic coherence since 
this particular term appears as off-diagonal element in the reduced density 
matrix of the atom. The time-evolution of population inversion shows a 
striking resemblance to the entanglement if the initial field state is a 
squeezed vacuum, see Figs. \ref{fig:pisv} and \ref{fig:entanglesv}. This 
feature is not to be seen in the case of coherent states, see 
Figs. \ref{fig:pics} and \ref{fig:entanglecs}. This observation is explained
 by the relationship, Eqn. (18), between the linear entropy and the 
population inversion. It turns out, as indicated by numerical calculations,
 that the atomic coherence vanishes during evolution if the initial state is
 a squeezed vacuum. Consequently, $L=1-W_{S}^{2}$ and there is a good 
resemblance of the $L(t)$ and $W_{S}(t)$ profiles. In the case of CS, the 
atomic coherence is not small. As a result, the  $L(t)$ and $W_{S}(t)$ 
profiles do not match. Interestingly, in this case the atomic coherence 
spikes to larger values during the course of evolution. Whenever, the 
coherence is large, the reduced density matrix is less mixed. This, in turn,
 means that the entanglement of the composite system is less. This is the 
origin of low entanglement spikes in the case of atom-coherent state 
interaction, refer Fig. \ref{fig:entanglecs}. \\
\noindent

           The time-evolution of linear entropy ($L$) is shown in 
Fig. \ref{fig:entanglecs}, when the cavity field is initially a coherent 
state. In Fig. \ref{fig:entanglesv}, results on the dynamics of $L$  are 
given for the squeezed vacuum case. The first row of the figures correspond
 to $\Delta=0$ (resonant case). The figures in the first column correspond  
to $k=0$ while those in the  second column correspond to $k=10^{-3}$. In the 
resonant case, the amplitude of variation of $L$ smaller if the value of $k$ 
is nonzero.  The effect of deformation is to minimize the effectiveness of the
 interaction in entangling the atom and the resonant field. This readily 
corroborates with the fact that the population inversion is nearly unity for 
this case, refer Fig. \ref{fig:pics} and \ref{fig:pisv}. Higher values of 
population inversion ($\approx 1 $) imply that the atomic state is 
$\approx |e\rangle$; consequently, the composite system is a product state 
and the linear entropy (equivalently, the entanglement) is nearly zero. Hence,
 with the increase of $k$, the entanglement in the atom-field system becomes 
zero, even in the resonant case. In the non-resonant NJC model, higher values 
of $k$ make the time-evolved atomic state to have substantial overlap with 
the ground state. This is in contrast with the non-resonant JC model in which 
the interaction becomes less effective in causing transitions as detuning 
increases. Thus, in the non-resonant case, the NJC interaction is more 
effective in entangling the atom and the field than the JC interaction. \\

           A simpler way of assessing the effectiveness of the interaction in
 entangling the field and the atom is to study the mean entanglement 
$\bar{L}$, defined as \cite{hou}
\begin{equation}
\bar{L}=\frac{1}{T}\int_0 ^T L(t')dt',
\end{equation}
is used.  The value of $T$ is chosen so that gT is 100. The values of the mean
 entanglement is shown in Table 1 for the single mode case.  Small nonzero 
values of $k$, of the order of $10^{-4}$ leads to better entanglement than 
zero or large values for $k$.  In the resonant case $\bar{L}$ approaches zero
 as $k$ increases from $10^{-4}$ to $10^{-3}$.  In the non-resonant case the 
entanglement does not become small even if $k$ increases to $10^{-3}$.  It is 
interesting to note that in the single-mode case, entanglement is more 
effective in the non-resonant case if the initial field is a CS in comparison 
to the case of initial field being in the SV.\\
\begin{table}
\caption{Mean entanglement in single-mode NJC model}
\begin{tabular}{|c|c|c|c|}
\hline
 \multicolumn{1}{|c|}{} & \multicolumn{1}{c|}{} & \multicolumn{2}{|c|}
{Mean entanglement} \\
\cline{3-4}
~~~$\Delta$~~~ & k & Coherent state & Squeezed vacuum \\ \hline
0 & 0 & 0.84 & 0.97 \\ 
0 & $ 1 \times 10^{-4} $ & 0.89 & 0.94 \\
0 & $ 1 \times 10^{-3} $ & 0.04 & 0.34 \\
0.01 & 0 & 0.56 & 0.64 \\
0.01 & $ 1 \times 10^{-4} $ & 0.91 & 0.77 \\
0.01 & $ 1 \times 10^{-3} $ & 0.75 & 0.26 \\
0.0161 & 0 & 0.3 & 0.44 \\
0.0161 & $ 1 \times 10^{-4} $ & 0.74 & 0.62 \\
0.0161 & $ 1 \times 10^{-3} $ & 0.81 & 0.24 \\
\hline
\end{tabular}
\end{table}
\\
\noindent
{\large\bf III. Two mode NJC model}\\

   In this section, the interaction between a two-level atom and a two-mode 
cavity field is studied. Two types of fields are considered: pair coherent 
(PC) state $|\zeta \rangle$ \cite{pcagarwal} and two-mode squeezed vacuum 
$|\mu \rangle $ (TSV) \cite{cavesschumaker}. The respective Fock state 
representations are: 
\begin{equation}
|\zeta \rangle =  N_{0}\sum_{n}  \,\frac{\zeta^{n}}{n!} |n,n\rangle,  
\end{equation}
where $N_{0}=1/\sqrt{I_{0}(2|\zeta|)} $, $I_{0}(2|\zeta|) $ is the modified
 Bessel function of order zero and $\zeta \in\it{C} $ is the amplitude 
of the state;
\begin{equation}
|\mu \rangle = \sqrt{1-|\mu|^2}\,\,\sum_{n} {\mu}^n \,\, |n,n\rangle,
\end{equation}
where $\mu \in \it{C}$ is the amplitude of state and $0 \le |\mu|<1$. Both
 these states involve superposition of paired states $|n,n\rangle$, which have
 equal number of quanta in both modes. The unpaired states 
$|n,m\rangle (n \ne m) $ are not present. In the quantum entanglement context,
 the TSV is an important example of entangled Gaussian state and the PC state
 is entangled but non-Gaussian.  Interestingly, TSV arises also in the 
context of teleportation of uniformly accelerated objects\cite{alsing}.
\\
\noindent

      The Hamiltonian of the system is taken to be $\hat{H}_{T}$, the 
two-mode extension of $\hat{H}_{S}$. The state of the system at time ``t" 
is $|\Psi(t) \rangle = \sum_{n=0}^{\infty}\,(C_{e,n,n}(t) |e,n,n\rangle 
+ C_{g,n,n}(t)|g,n,n\rangle) $. Unpaired states are absent in the evolved 
state $|\Psi(t)\rangle$ as well. This feature is due to the type of 
interaction being considered and the absence of unpaired states in the initial
 state. In what follows, we set $k_{1}=k_{2}=k$. With this choice for the two
 deformation parameters, the interaction term 
$\hat{H}_{I} = \lambda \sqrt{(1+k\hat{a}^{\dagger}\hat{a})
(1+k\hat{b}^{\dagger}\hat{b})}\hat{a}\hat{b}\hat{\sigma}_{+}+\hbox{adjoint}$
 gives,
\begin{eqnarray}
\hat{H}_{I}|e,n,n\rangle &=& \lambda (1+kn+k)|g,n+1,n+1\rangle, \\ 
\hat{H}_{I}|g,n,n\rangle &=& \lambda (1+kn-k)|e,n-1,n-1\rangle .
\end{eqnarray}
That is, a paired state is transformed to another paired state. More 
precisely, under the action of $\hat{H}_{T}$, the paired state 
$|e,n,n\rangle$ becomes the paired state $|g,n+1,n+1\rangle$ and vice-versa.
 Hence, the evolved state is a superposition of paired states if the initial
 state is a superposition of paired states. Another consequence is that there
 are invariant subspaces in the Hilbert space of the tripartite system. It is
 easily recognized from the transformations indicated in Eqs. (21) and (22), 
that the span of $\{|e,n,n \rangle,  |g,n+1,n+1\rangle\} $, for a fixed $n$, 
is invariant under the action of $\hat{H}_{I}$. Moreover, the invariant 
subspaces are disjoint. These features imply that the dynamics of the 
coefficients $C_{e,n,n}(t)$ and $C_{g,n+1,n+1}(t)$ are coupled to each other 
and do not depend on the dynamics of other coefficients. The Schrodinger 
equation for the dynamics of the coefficients $C_{e,n,n}(t)$ and 
$C_{g,n+1,n+1}(t)$ is the matrix equation, 
\begin{equation}
 i \frac{d}{dt}
\left(
\begin{array}{c}
   C_{e,n,n}(t)    \\
   C_{g,n+1,n+1}(t)
\end{array}
\right)
=\left[
\begin{array}{cc}
h_{11} & h_{12} \\
h_{21} & h_{22}
\end{array}
\right]
\left(
\begin{array}{c}
C_{e,n,n}(t) \\
C_{g,n+1,n+1}(t)
\end{array}
\right)
.
\end{equation}
Here  $h_{11} = \omega_{1}(n+\frac{1}{2}+kn^{2}-kn) 
+ \omega_{2}(n+\frac{1}{2}+kn^{2}-kn) + \frac{\Delta}{2}$, \\ 
$h_{22} = \omega_{1}(n+\frac{1}{2}+kn^{2}+kn) 
+ \omega_{2}(n+\frac{1}{2}+kn^{2}+kn) - \frac{\Delta}{2}$, \\
and $h_{12} = h_{21} = \lambda (1+n)(1+kn)$ . Here, 
$\Delta = \nu -(\omega_{1}+\omega_{2})$ is the detuning. This matrix equation
 is solved to give 
\begin{eqnarray}
e^{-i\Delta_{n,n}t/2} C_{e,n,n} (t) &=&[\cos(\frac{\Omega_{n,n}t}{2}) 
- \frac{i\Delta_{n,n}}{\Omega_{n,n}}\sin(\frac{\Omega_{n,n}t}{2})]C_{e,n,n}(0)
 \nonumber\\ & &  -\frac{2i \lambda \eta_{n,n}}{\Omega_{n,n}}\,\, 
\sin(\frac{\Omega_{n,n}t}{2})\,\, C_{g,n+1,n+1}(0),
\end{eqnarray}
\begin{eqnarray}
e^{i\Delta_{n,n}t/2} C_{g,n+1,n+1} (t) &=& [\cos(\frac{\Omega_{n,n}t}{2}) 
+ \frac{i\Delta_{n,n}}{\Omega_{n,n}}\sin(\frac{\Omega_{n,n}t}{2})]
C_{g,n+1,n+1}(0)\nonumber\\
       & &-\frac{2i \lambda \eta_{n,n}}{\Omega_{n,n}}\,\, 
\sin(\frac{\Omega_{n,n}t}{2})\,\, C_{e,n,n}(0),
\end{eqnarray}
where $\Delta_{n,n} = \Delta - 2k\omega_{1} n - 2k\omega_{2} n , \,\,
\Omega_{n,n} = \sqrt{\Delta_{n,n}^2 + 4 {\lambda}^{2}\eta_{n,n}^{2}}$ and 
$\eta_{n,n} = (1+n)(1+kn)$. These analytical solutions are used in calculating
 population inversion and entanglement measures in the tripartite system. \\

\noindent

\begin{figure}
\centering
\includegraphics[height=15cm,width=15cm]{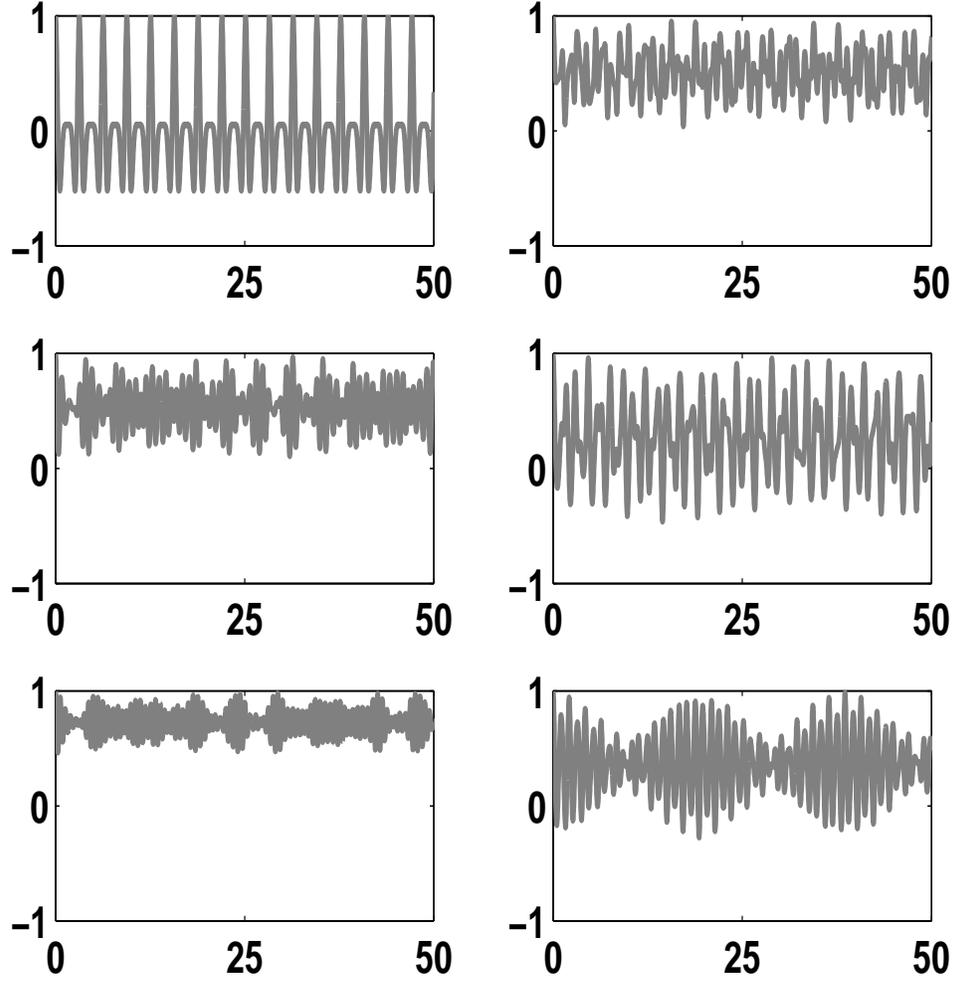}
\caption {{\small Variation of population inversion with scaled time 
($\lambda t$) when the initial state is $|\zeta\rangle|e\rangle$; the field in
 the pair coherent state $|\zeta\rangle$ ($\zeta = 1.778$) and the atom in the
 excited state $|e\rangle$. The coupling constant is $\lambda = 0.002$, the 
mean number of photons in the field is $\bar{N}=3$ and the critical detuning 
is $\Delta^{\prime}_{c} = 0.016$. Successive rows correspond to $\Delta=0,\,
\Delta=0.01( <\Delta^{\prime}_{c}),\,\Delta=\Delta^{\prime}_{c}$. Successive 
columns correspond to $k$ equal to $0,\,2 \times 10^{-3}$.} }
\label{fig:pipc}
\end{figure}

\begin{figure}
\centering
\includegraphics[height=15cm,width=15cm]{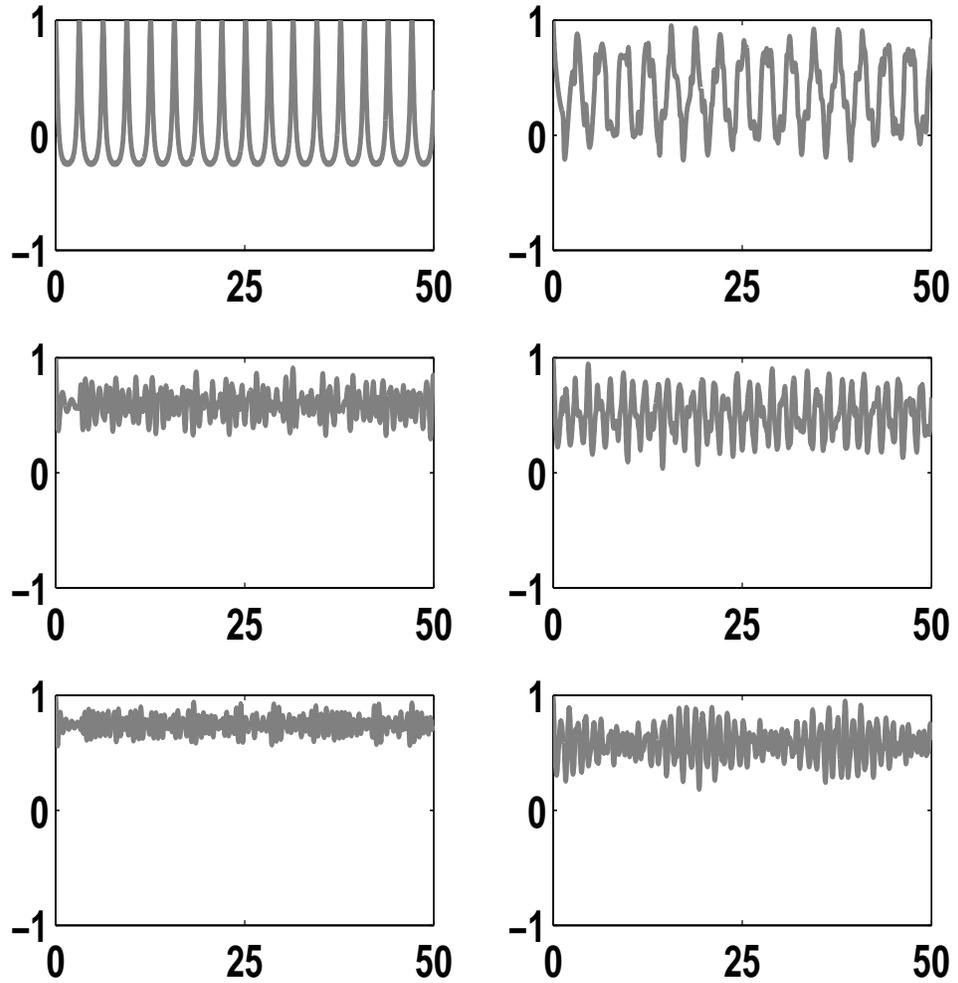}
\caption {{\small Variation of population inversion with scaled time 
($\lambda t$) when the initial state is $|\mu \rangle|e\rangle$; the field 
state in the two-mode squeezed vacuum state $|\mu \rangle$ ($\mu=1.032$) and
 the atom in the excited state $|e\rangle$. Successive rows correspond to 
$\Delta=0,\,\Delta=0.01( <\Delta^{\prime}_{c}),\,\Delta=\Delta^{\prime}_{c}$.
 Successive columns correspond to $k$ equal to $0,\,2 \times 10^{-3}$. Other 
parameters are the same as in Fig. 6.}}
\label{fig:pitsv}
\end{figure}

      The reduced density operator $\hat{\rho}_{A}$ for the atomic system is 
\begin{equation}
\hat{\rho}_{A} =
\left(
\begin{array}{cc}
\sum_{n}|C_{e,n,n}(t)|^{2}   & \sum_{n}C_{e,n,n}^{\ast}(t)C_{g,n,n}(t) \\
\sum_{n}C_{e,n,n}(t)C_{g,n,n}^{\ast}(t) & \sum_{n}|C_{g,n,n}(t)|^{2}
\end{array}
\right)
.
\end{equation}
For the two-mode case, the population inversion, denoted by $W_{T}(t)$, is 
given by
\begin{equation}
W_{T}(t) = \sum_{n}[ |C_{e,n,n}(t)|^{2} - |C_{g,n,n}(t)|^{2}].
\end{equation}
The evolution of population inversion is given in Fig. \ref{fig:pipc} for the 
PC state and in Fig. \ref{fig:pitsv} for the TSV. In both the figures, the 
successive columns correspond to $k=0$ and $k=2 \times 10^{-3}$ respectively. 
And  the successive rows correspond to different detuning parameter values 
($\Delta$) starting from zero detuning. In the resonant two-mode JC model, 
\textit{i.e.,} $\Delta=0$ and $k = 0$, the Rabi frequencies satisfy 
$\Omega_{n,n}t=gt(1+n)$; these frequencies are commensurate and so the 
dynamics is periodic. As a consequence, $W_{T}(t)$  shows periodic features 
in the resonant two-photon JC model. At the time instances when $gt=l\pi$ 
($l$, a positive integer) the atom returns to the initial state, unlike in the
 single mode case. This periodicity is destroyed in the non-resonant 
interaction as the Rabi frequencies are incommensurate. The atomic evolution 
from the excited state is increasingly inhibited with larger detuning which 
is evident from the column 1 of Figs. \ref{fig:pipc} and \ref{fig:pitsv}. It 
is to be noted that the Rabi frequency $\Omega_{n,n}$ depends only on a single
 quantum number ``$n$'', a consequence of the initial field being a 
superposition of paired state. As a function of $n$, the two-mode Rabi 
frequency exhibits a minimum at $n=\bar{n}$, obtained by solving 
$\frac{\partial \Omega_{n,n} }{\partial t} |_{n=\bar{n}}=0$, provided 
$k \ne 0$. The detuning corresponding to $n=\bar{n}$ is called critical 
detuning, which is given as

\begin{equation}
\Delta_{\bar{n},\bar{n}} = \Delta^{\prime}_{c} =  k\omega \bar{N} 
+ \frac{g^2}{k\omega}[(1+k)(2+\bar{N}+k \bar{N}
+\frac{3}{2}\bar{N}^{2} k)+\frac{1}{2}\bar{N}^{3}k^{2}] ,
\end{equation}
where $\bar{N}=2\bar{n}$ is the total mean photon number of the two-mode 
radiation field. At critical detuning, the evolution of $W_{T}(t)$ has 
noticeable features. The envelope of $W_{T}(t)$ appears periodic and this is
 lost when detuning shifts away from $\Delta^{\prime}_{c}$. This is seen by
 comparing the figures in the third row($\Delta=\Delta^{\prime}_{c}$) with 
the those in the second row($\Delta < \Delta^{\prime}_{c}$) of 
Fig. \ref{fig:pipc}. Further, nonzero $k$ value (refer the second column of 
Fig. 6) significantly influences the envelope pattern in the population 
inversion. As in the case of single mode NJC, the critical detuning alters 
the population inversion dynamics qualitatively. A difference in the 
evolution at critical detuning, in the single-mode and two-mode cases is to 
be pointed out. In the single-mode case, super structures are seen if the 
initial field state is a CS while such structures are not prominent if the 
initial state is a SV. But in the two-mode case, similar structures in the 
envelope of $W_{T}(t)$ are observed whether the initial field state is a PC 
state or a TSV.  \\

\noindent

\begin{figure}
\centering
\includegraphics[height=15cm,width=15cm]{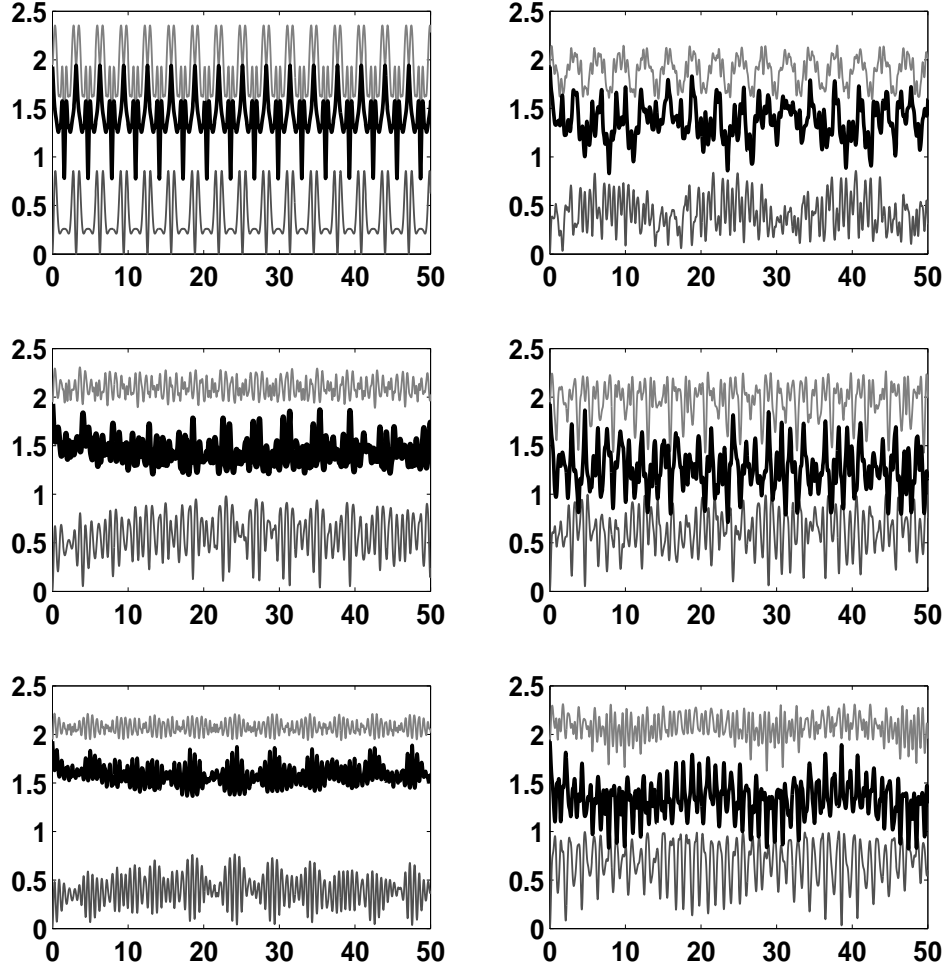}
\caption {{\small Evolution of various biparitite entanglement measures versus
 scaled time ($\lambda t$). The initial state of the field is the pair 
coherent state $|\zeta\rangle$ ($\zeta = 1.778$) and that of the atom is the
 excited state $|e\rangle$. The first column corresponds to $k=0$ and the 
second column corresponds to $k=2\times 10^{-3}$. The top most row corresponds
 to $\Delta=0$ and the successive rows correspond to $\Delta=0.01$ and 
$\Delta=0.0161$. In each figures, the quantities shown are 
$T_{A,F_{1}\otimes F_{2}}$(bottom), $E(\rho_{F_{1},F_{2}})$(middle) and 
$T_{A\otimes F_{1},F_{2}}$(top). Other parameters are the same as in Fig. 6.}}
\label{fig:entanglepc}
\end{figure}

\begin{figure}
\centering
\includegraphics[height=15cm,width=15cm]{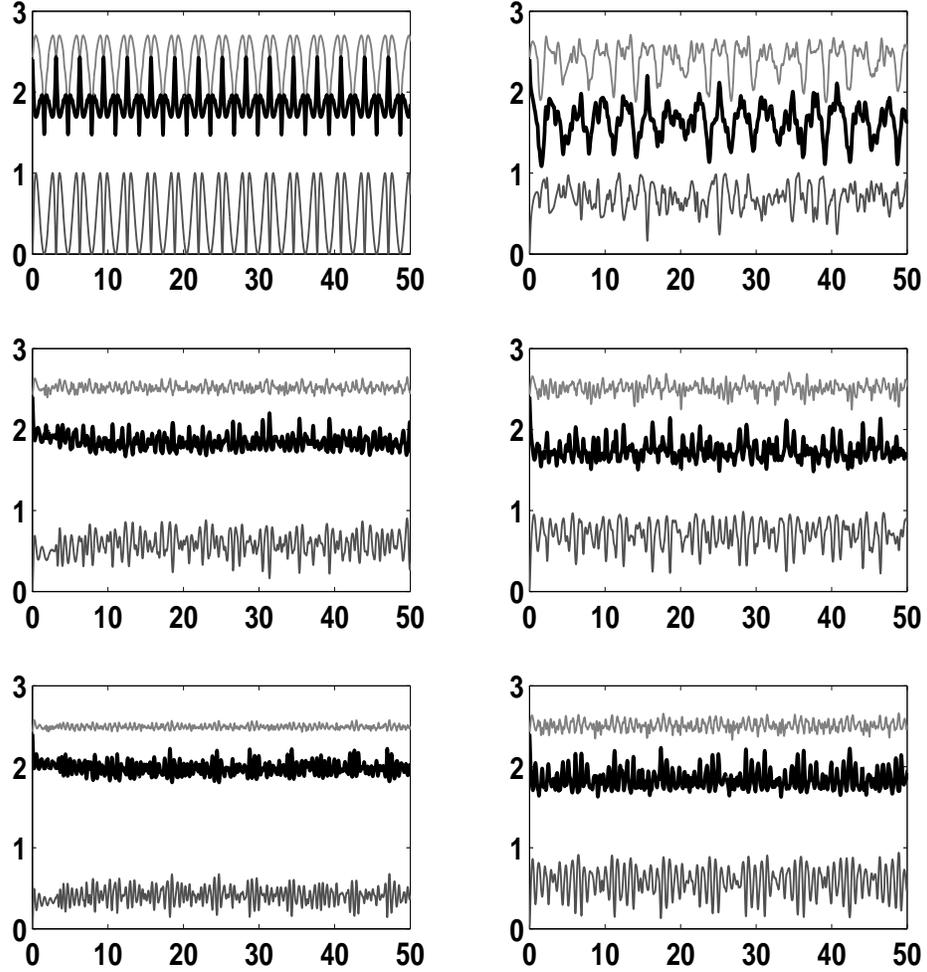}
\caption {{\small Evolution of various biparitite entanglement measures with 
time ($\lambda t$). The initial state of the field is the two-mode squeezed 
vacuum state $|\mu \rangle$ ($\mu =1.032$) and that of the atom is the 
excited state $|e\rangle$. The first column corresponds to $k=0$ and the 
second column corresponds to $k=2\times 10^{-3}$. The top most row corresponds
 to $\Delta=0$ and the successive rows correspond to $\Delta=0.01$ and 
$\Delta=0.0161$. In each figures, the quantities shown are 
$T_{A,F_{1}\otimes F_{2}}$(bottom), $E(\rho_{F_{1},F_{2}})$(middle) and 
$T_{A\otimes F_{1},F_{2}}$(top). Other parameters are the same as in Fig. 6.}}
\label{fig:entangletsv}
\end{figure}

      The system described by $\hat{H}_{T}$ is a tripartite system: a 
two-level atom and two field modes. The Hilbert spaces associated with the 
field modes are of infinite dimension. There is no known measure of 
entanglement if any subsystem of a tripartite system is of infinite 
dimension. An assessment of entanglement  in the system is obtained by 
studying bipartite entanglement in different partitions of the system. In the 
present case, the relevant partitions are: $(\hbox{i})$ two-level atom as a 
subsystem and the two field modes together form the other subsystem; 
$(\hbox{ii})$ two-level atom along with one of the field modes as one 
subsystem and the other field mode as the other subsystem; and 
$(\hbox{iii})$ the third possibility is to exchange the roles of the fields 
in the previous case. The entanglement measures of the different bipartite 
partitions are called tangles ($T$). The states considered in this work are 
pure states and hence the various tangles are chosen to be the respective 
von Neumann or linear entropies. Thus,   
\begin{equation}
T_{A,F_{1}\otimes F_{2}} = 2[1-Tr(\hat{\rho}_{A}^{2})] =  
2[1-Tr(\hat{\rho} _{F_{1}\otimes F_{2}}^{2})],\end{equation}
\begin{equation}
T_{A \otimes F_{1} , F_{2}} = -Tr[\hat{\rho}_{A\otimes F_{1}}\,\, 
\log_{2}(\hat{\rho}_{A\otimes F_{1}})] =  -Tr[\hat{\rho}_{F_{2}}\,\, 
\log_{2}(\hat{\rho}_{F_{2}})],
\end{equation}
\begin{equation}
T_{A \otimes F_{2} , F_{1}} = -Tr[\hat{\rho}_{A\otimes F_{2}}\,\, 
\log_{2}(\hat{\rho}_{A\otimes F_{2}})] =  -Tr[\hat{\rho}_{F_{1}}\,\, 
\log_{2}(\hat{\rho}_{F_{1}})].
\end{equation}
The suffix ($A, F_{1}\otimes F_{2}$) indicates that the atom is treated as a 
subsystem and the two field modes ($F_{1}$ and $F_{2}$) are treated as a 
single entity is another subsystem of the relevant partition. Similar 
convention is adapted for the other suffixes. The initial field state, a 
superposition of paired states is symmetric under the exchange of photon 
numbers of the two modes and this symmetry is preserved during evolution. 
Consequently, $T_{A\otimes F_{1} , F_{2}} $ and $T_{A\otimes F_{2} , F_{1}}$ 
 are identical. In Figs. \ref{fig:entanglepc} and \ref{fig:entangletsv}, 
only $T_{A\otimes F_{1} , F_{2}} $ is shown. The calculation of 
$T_{A\otimes F_{1} , F_{2}}$ requires the reduced density matrix of the atom 
and the first mode, which is   
\begin{eqnarray}
\hat{\rho}_{A\otimes F_{1}} &=& \sum_{n}\vert C_{e,n,n}\vert^{2}
\vert e n\rangle\langle
 e n\vert
+C_{e,n,n}C^{\ast}_{g,n,n}|e n \rangle\langle g n|\nonumber\\ 
                     & &+C_{g,n,n}C^{\ast}_{e,n,n}|g n \rangle\langle e n|
+|C_{g,n,n}|^{2}|g n \rangle\langle g n\vert.
\end{eqnarray}

\noindent

              Another quantity that is of interest is the entanglement 
between the two field modes. Quantum relative entropy 
$E_{F_{1},F_{2}}$ is a measure of entanglement in a bipartite 
field state involving only the paired states \cite{rains}. The field density
 matrix is
\begin{eqnarray}
\hat{\rho}_{F_{1},F_{2}} &=& \sum_{r=e,g} \langle r |\Psi(t) \rangle 
\langle \Psi(t) | r \rangle\\
            &=&  \sum_{n,m} a_{n,m}  \left|n,n \rangle \langle  m,m\right|,
\end{eqnarray}
obtained by tracing over the atomic states in the complete density matrix 
of the tripartite system. The density matrix element $a_{n,m}$, in terms of 
the coefficients $C_{e,n,n}$ and $C_{g,n,n}$, is 
$C_{e,n,n}C^{\ast}_{e,m,m}+C_{g,n,n}C^{\ast}_{g,m,m}$. The von Neumann 
entropy is
\begin{equation}
S(\hat{\rho}_{F_{1},F_{2}}) = -Tr[\hat{\rho}_{F_{1},F_{2}}\,\, 
\log_{2}(\hat{\rho}_{F_{1},F_{2}})],
\end{equation}
and the quantum relative entropy is
\begin{equation}
E_{F_{1},F_{2}} = -\sum_{n} a_{n,n} \log_{2}(a_{n,n}) - 
S(\hat{\rho}_{F_{1},F_{2}}).
\end{equation} \\
 
\noindent

     The tangles $T_{A, F_{1}\otimes F_{2}}$, $T_{A\otimes F_{1},F_{2}}$ and 
the relative entropy $E(\hat{\rho}_{F_{1},F_{2}})$ are shown in 
Figs. \ref{fig:entanglepc} and \ref{fig:entangletsv} for the system when the 
initial states are PC and TSV respectively. As in the case of 
Figs. \ref{fig:pipc} and \ref{fig:pitsv}, the two columns correspond to $k=0$ 
and $2 \times 10^{-3}$ respectively. The three rows of figures correspond to 
$\Delta=0$, $\Delta=0.01 (<\Delta^{\prime}_{c})$, and 
$\Delta = 0.0161(=\Delta^{\prime}_{c})$ respectively. The mean photon number 
of the two modes is taken to be 3. This ensures that the contribution from 
higher photon number states are negligible and numerical computations are 
reliable. Firstly, the time evolution profile of the tangle 
$T_{A\otimes F_{1},F_{2}}$ and the relative entropy 
$E_{F_{1},F_{2}}$ do not resemble that of the population 
inversion $W_{T}(t)$. However, $T_{A,F_{1}\otimes F_{2}}$ which is linear 
entropy, is related to $W_{T}(t)$ through
\begin{equation}
W_{T}^{2} = 1 - T_{A,F_{1}\otimes F_{2}}- 4|\sum C_{e,n,n}^{\ast}
C_{g,n,n}|^{2}.
\end{equation}
In spite of this relation, the profiles of $W_{T}$ and  
$T_{A,F_{1}\otimes F_{2}}$ are not similar. This is in contrast with the 
single-mode case wherein the profiles of the entanglement resemble with 
$W_{S}(t)$, at least for the smaller values of the parameter $k$. The reason 
for this dissimilarity is that in the two-mode case, the atomic coherence 
$|\sum C_{e,n,n}^{\ast}C_{g,n,n}|^{2}$ is comparable to 
$T_{A,F_{1}\otimes F_{2}}$ and does not become small either for the pair 
coherent state or the two-mode squeezed vacuum. \\
\noindent

          The profiles of the tangles and the relative entropy are periodic, 
when $k=0$ and $\Delta=0$, corresponding to the resonant two-mode JC model. 
When the deformation parameter $k$ is nonzero but small ($=2 \times 10^{-3}$), 
the periodicity in the evolution is destroyed. As noted earlier, this feature 
is present in the evolution of $W_{T}$ too. Another interesting aspect is the 
negative correlation between the time-evolutions of the tangle 
$T_{A,F_{1}\otimes F_{2}}$ and the relative entropy $E$, \textit{i.e.,} 
increase (decrease) of $T_{A,F_{1}\otimes F_{2}}$ is associated with decrease 
(increase) of $E$. The negative correlation may be interpreted as 
redistribution of entanglement, from the field-field entanglement to that 
between the  atom and the two modes. This feature is retained even if $k$ is 
nonzero implying that the nonlinearity and the intensity-dependent interaction
 do not hinder the redistribution of entanglement.\\
\noindent

            In the non-resonant case, the atom does not interact with the 
field effectively and hence the initial entanglement is not expected to vary
 much. It is essential to study the cases corresponding to 
$\Delta = \Delta^{\prime}_{c}$ and  $\Delta \ne \Delta^{\prime}_{c}$. There 
is an increase of amplitude in the tangle oscillation as $k$ increases to 
$2 \times 10^{-3}$ from zero as if the atom-field interaction is effective as 
in the resonant case. Thus, small non-zero $k$ leads to a ``resonance" effect
 in the non-resonant situation. At slightly larger values of 
$k\approx 10^{-2}$, the quantity $T_{A, F_{1}\otimes F_{2}}$ becomes 
insignificant indicating very less entanglement between the atom and the two
 modes. However, this fact is to be expected; for such values of $k$ the 
population inversion $W_{T}(t)$ is nearly unity implying that the atomic 
state is $\sim|e\rangle$ and hence the atom is not entangled with the fields.
 The need for larger $k$, in comparison to $k=10^{-3}$ in the single-mode 
case, to disentangle the atom and the two fields is because of the smaller 
photon number ($\approx 3$) in the two-mode case.\\

           As in the single mode case, it is useful to define the mean of 
the tangles and the relative entropy to characterize the effectiveness of 
the interaction in entangling the systems.  Definitions similar to $\bar{L}$ 
are adopted for the entanglement measures  $T_{A,F_{1}\otimes F_{2}}$ and 
$T_{A\otimes F_{1},F_{2}}$ and the relative entropy $E_{F_{1}, F_{2}}$. The 
mean is computed over a period $T$ such that $gT=20$.  From the table it seen
 that there is not as much effect on the mean values with increase of $k$ when
 compared with the single mode case.\\
\begin{table}[htp]
\caption{Mean tangles and Mean relative entropy in two-mode NJC model}
\begin{tabular}{|c|c|c|c|c|c|c|c|}
\hline
\multicolumn{1}{|c|}{} &\multicolumn{1}{c|}{} & \multicolumn{2}{c|} 
{Mean $T_{A,F_{1}\otimes F_{2}}$ } & \multicolumn{2}{c|}
{Mean $T_{A\otimes F_{1}, F_{2}}$} & \multicolumn{2}{c|}
{Mean $E_{F_{1}, F_{2}}$}\\
\cline{3-8}
~~~$\Delta$~~~ & k &~~PC~~ & TSV & ~~PC~~ & TSV & ~~PC~~ & TSV \\ \hline
 0 & 0 & 0.39 & 0.50 & 1.89 & 2.41 & 1.42 & 1.86 \\
 0 & $2 \times 10^{-3}$ & 0.41 & 0.7 & 1.88 & 2.4 & 1.38 & 1.65 \\
 0.01 & 0 & 0.58 & 0.58 & 2.1 & 2.52 & 1.45 & 1.86 \\
 0.01 & $2 \times 10^{-3}$ & 0.61 & 0.7 & 1.97 & 2.51 & 1.27 & 1.74 \\
 0.0161 & 0 & 0.40 & 0.40 & 2.08 & 2.49 & 1.60 & 1.98 \\
 0.0161 & $2 \times 10^{-3}$ & 0.66 & 0.59 & 2.09 & 2.51 & 1.33 & 1.84 \\
\hline
\end{tabular}
\end{table}
\\          
{\large\bf IV. Summary}\\

\noindent
      Nonlinearity and intensity-dependent interactions affect entanglement 
between an atom and a cavity field. The time-evolution due to $\hat{H}_{S}$,
  wherein both nonlinearity (Kerr type) and intensity-dependent interaction 
($\propto \sqrt{1+\hbox{intensity}}$) are included, is exactly solvable. The 
dynamical symmetry of the Hamiltonian is a deformed Heisenberg-Weyl algebra. 
When the deformation vanishes, the algebra is the Heisenberg-Weyl algebra. 
The deformation can be continuously varied  so that the algebra is SU(1,1) 
when the deforming parameter is unity. The deforming parameter has the 
physical significance  that the nonlinearity and the interaction are 
proportional to it. The Hamiltonians considered in this work approximate 
many other Hamiltonians that are useful in modeling various quantum optical 
systems.  \\

\noindent

          The deformed Hamiltonian $\hat{H}_{S}$ makes the Rabi frequency to 
attain a minimum value which allows for critical detuning. When the cavity 
field is chosen so that the mean photon number is $\bar{n}$ and there is 
critical detuning between the field and the atom, the dynamics exhibits 
``super structures" in the evolution of population inversion. In the case of 
single-mode interaction, with a small deformation ($k \approx 10^{-4}$) the
 evolution of population inversion and entanglement are very different from 
those in the usual JC interaction($k=0$). One of the effects of nonzero $k$ 
is that the collapses and revivals in the population inversion are less 
pronounced due to the Kerr nonlinearity. In reality, weak nonlinearities and 
intensity-dependent interactions are often unavoidable. This implies that in 
experiments it is really hard to see as many distinct  
``revivals and collapses" as indicated by the usual single-mode JC model. At 
resonance, the effect of higher $k$ is to inhibit the evolution of the atom 
from its excited state. These features are common to both the single-mode and 
the two-mode NJC models. Some aspects of population inversion dynamics are 
different for the single-mode and two-mode models. In the undeformed, resonant
 case the population inversion $W_{S}(t)$ exhibits `collapse-revival' 
structure when the initial state is a coherent state and these features are 
absent if the initial field is a squeezed vacuum. In the two-mode case, such 
structures are seen when the cavity is either a pair coherent state or a 
two-mode squeezed vacuum. In the nonresonant case, the deformed Hamiltonian 
generates more entanglement than the undeformed case. In the resonant case, 
small deformations tend to reduce the effectiveness of the interaction and 
hence the entanglement is reduced. In the two-mode case, the tangle 
$T_{A,F_{1}\otimes F_{2}}$ and the relative entropy $E_{F_{1}, F_{2}}$ of the 
two fields exhibit 
negative correlation in their dynamics. This is interpreted as the 
redistribution of entanglement from that between the  fields to that between 
the  atom and the fields and vice versa.\\

\noindent
\begin{center}
{\large\bf Acknowledgment}
\end{center}

   The authors acknowledge Dr. M. V. Satyanarayana for useful discussions. \\

%\noindent
\centering

%%%%%%%%%%%%%%%%%%%%%%%%%%%%%%%%%%%%%%%%%%%%%%%%%%%%%%%%%%%%%%%%%%%%%%%%%%%%%%%
\newpage              %%%%%%   FIGURE 1  %%%%%%%   
%%%%%%%%%%%%%%%%%%%%%%%%%%%%%%%%%%%%%%%%%%%%%%%%%%%%%%%%%%%%%%%%%%%%%%%%%%%%%%%
%%%%%%%%%%%%%%%%%%%%%%%%%%%%%%%%%%%%%%%%%%%%%%%%%%%%%%%%%%%%%%%%%%%%%%%%%%%%%%
\newpage              %%%%%%%%%%% FIGURE 2  %%%%%%%
%%%%%%%%%%%%%%%%%%%%%%%%%%%%%%%%%%%%%%%%%%%%%%%%%%%%%%%%%%%%%%%%%%%%%%%%%%%%%%
%%%%%%%%%%%%%%%%%%%%%%%%%%%%%%%%%%%%%%%%%%%%%%%%%%%%%%%%%%%%%%%%%%%%%%%%%%%%%%%
\newpage     %%%%%%%%%%%%%  FIGURE 3 %%%%%%%%%%%
%%%%%%%%%%%%%%%%%%%%%%%%%%%%%%%%%%%%%%%%%%%%%%%%%%%%%%%%%%%%%%%%%%%%%%%%%%%%%%
%%%%%%%%%%%%%%%%%%%%%%%%%%%%%%%%%%%%%%%%%%%%%%%%%%%%%%%%%%%%%%%%%%%%%%%%%%%%%%%%
\newpage    %%%%%%%%%%%%%%% FIGURE 4 %%%%%%%%%%%%%%
%%%%%%%%%%%%%%%%%%%%%%%%%%%%%%%%%%%%%%%%%%%%%%%%%%%%%%%%%%%%%%%%%%%%%%%%%%%%%%%%
%%%%%%%%%%%%%%%%%%%%%%%%%%%%%%%%%%%%%%%%%%%%%%%%%%%%%%%%%%%%%%%%%%%%%%%%%%%%%%%
\newpage   %%%%%%%%%%%%%  FIGURE 5 %%%%%%%%%%%%%%%
%%%%%%%%%%%%%%%%%%%%%%%%%%%%%%%%%%%%%%%%%%%%%%%%%%%%%%%%%%%%%%%%%%%%%%%%%%%%%%%
%%%%%%%%%%%%%%%%%%%%%%%%%%%%%%%%%%%%%%%%%%%%%%%%%%%%%%%%%%%%%%%%%%%%%%%%%%%%%%%
\newpage    %%%%%%%%%%%%%%%%%% FIGURE 6 %%%%%%%%%%%%%%%%%%
%%%%%%%%%%%%%%%%%%%%%%%%%%%%%%%%%%%%%%%%%%%%%%%%%%%%%%%%%%%%%%%%%%%%%%%%%%%%%%%
%%%%%%%%%%%%%%%%%%%%%%%%%%%%%%%%%%%%%%%%%%%%%%%%%%%%%%%%%%%%%%%%%%%%%%%%%%%%%%%
\newpage   %%%%%%%%%%%%% FIGURE 7 %%%%%%%%%%%%%%%%%
%%%%%%%%%%%%%%%%%%%%%%%%%%%%%%%%%%%%%%%%%%%%%%%%%%%%%%%%%%%%%%%%%%%%%%%%%%%%%%%%
%%%%%%%%%%%%%%%%%%%%%%%%%%%%%%%%%%%%%%%%%%%%%%%%%%%%%%%%%%%%%%%%%%%%%%%%%%%%%%%
\newpage   %%%%%%%%%%%%%%%%%% FIGURE 8 %%%%%%%%%%%%%%
%%%%%%%%%%%%%%%%%%%%%%%%%%%%%%%%%%%%%%%%%%%%%%%%%%%%%%%%%%%%%%%%%%%%%%%%%%%%%%%
%%%%%%%%%%%%%%%%%%%%%%%%%%%%%%%%%%%%%%%%%%%%%%%%%%%%%%%%%%%%%%%%%%%%%%%%%%%%%%%
\newpage    %%%%%%%%%%%%%%%%%% FIGURE 9 %%%%%%%%%%%%%%%%
%%%%%%%%%%%%%%%%%%%%%%%%%%%%%%%%%%%%%%%%%%%%%%%%%%%%%%%%%%%%%%%%%%%%%%%%%%%%%%
%%%%%%%%%%%%%%%%%%%%%%%%%%%%%%%%%%%%%%%%%%%%%%%%%%%%%%%%%%%%%%%%%%%%%%%%%%%%%%
\end{document}